\documentclass[journal]{IEEEtran}
\ifCLASSINFOpdf
\else
\fi
\usepackage{amsmath}        
\usepackage{graphicx}
\usepackage{color}
\usepackage{enumitem} 
\usepackage{multirow}

\DeclareMathOperator*{\argmin}{arg\,min}

\newcommand{\measurement}{\mathbf{y}}
\newcommand{\signal}{\mathbf{x}}
\newcommand{\reps}{\mathbf{c}} 
\newcommand{\repsHat}{\hat{\reps}} 

\begin{document}
\title{Deep Learning of Compressed Sensing Operators with Structural Similarity Loss}
%
%
%
\author{Y. Zur and
	A. Adler
	\thanks{Yochai Zur is with the Department of Computer-Science, Technion Israel Institute of Technology.}
	\thanks{Amir Adler is with the McGovern Institute for Brain Research, Massachusetts Institute of Technology.}
}

%

\maketitle

\begin{abstract}
Compressed sensing (CS) is a signal processing framework for efficiently reconstructing a signal from a small number of measurements, obtained by linear projections of the signal. In this paper we present an end-to-end deep learning approach for CS, in which a fully-connected network performs both the linear sensing and non-linear reconstruction stages. During the training phase, the sensing matrix and the non-linear reconstruction operator are \emph{jointly} optimized using \emph{Structural similarity index (SSIM)} as loss rather than the standard Mean Squared Error (MSE) loss. We compare the proposed approach with state-of-the-art in terms of reconstruction quality under both losses, i.e. SSIM score and MSE score.
\end{abstract}


%
\IEEEpeerreviewmaketitle

\section{Introduction}
Compressed sensing (CS) \cite{donoho2006compressed,candes2008introduction} is a mathematical framework that defines the conditions and tools for the recovery of a signal from a small number of its linear projections (i.e. measurements). In the CS framework, the measurement device acquires the signal in the linear projections domain, and the full signal is reconstructed by convex optimization techniques. CS has diverse applications including image acquisition \cite{romberg2008imaging}, radar imaging \cite{potter2010sparsity}, Magnetic Resonance Imaging (MRI) \cite{lustig2008compressed,murphy2012fast}, spectrum sensing \cite{axell2012spectrum}, indoor positioning \cite{feng2012received}, bio-signals acquisition \cite{dixon2012compressed}, and sensor networks \cite{li2013compressed}. In this paper we address the CS problem by using a novel loss function, the SSIM loss. Our approach is based on a deep neural network \cite{bengio2009learning}, which simultaneously learns the linear sensing matrix and the non-linear reconstruction operator under the SSIM loss.

The contributions of this paper are two-fold: (1) It presents for the first time, to the best knowledge of the authors, a training with SSIM loss fonction of a deep neural network for the tasks of reconstruction; and (2) During training, the proposed network \emph{jointly} optimizes both the linear sensing matrix and the non-linear reconstruction operator.

This paper is organized as follows: section \ref{Compressed Sensing} reviews compressed sensing concepts. Section \ref{SSIM} reviews the novel SSIM loss function. Section \ref{The Proposed Approach} presents the end-to-end deep learning approach. Section \ref{Results} discusses structure and training aspects, while evaluating the performance of the proposed approach for image reconstruction, and comparing it with state-of-the-art alternatives. Section \ref{Conclusions} concludes the paper and discusses future research directions.

\begin{figure*}[h]
	\includegraphics[width=\linewidth,height=7cm]{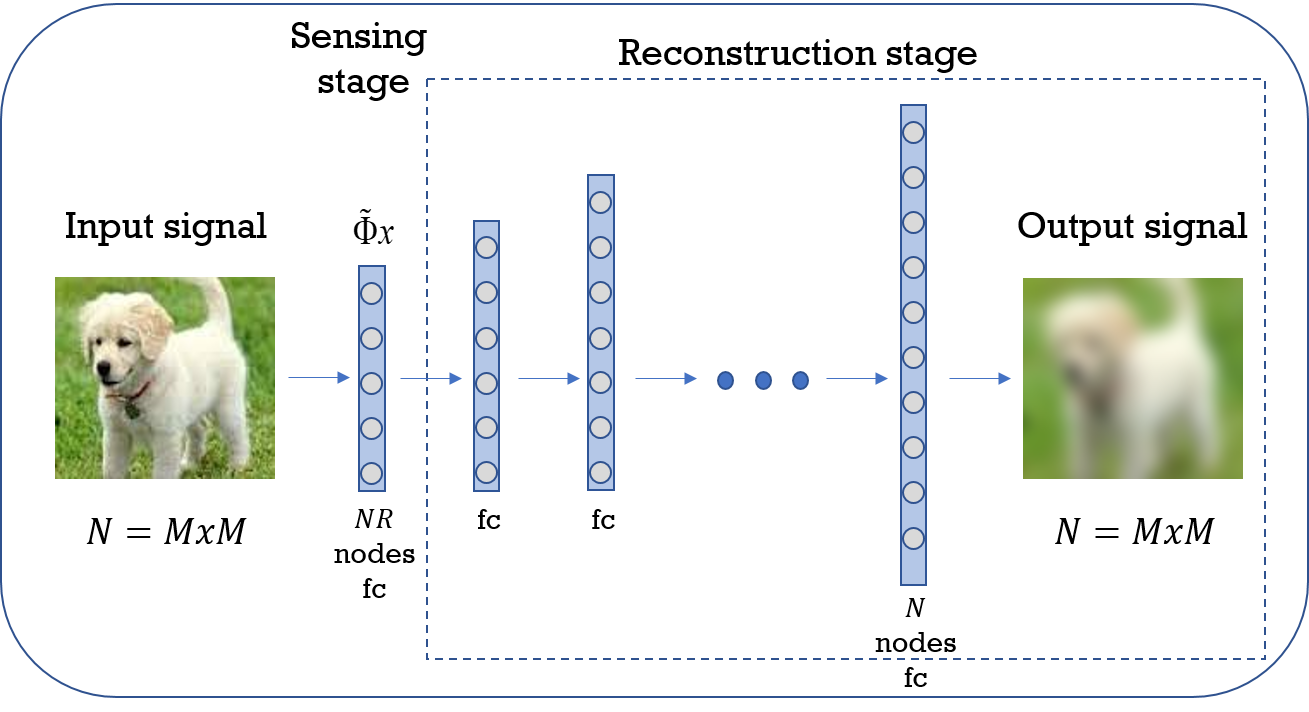}
	\caption{Scheme of the end-to-end deep neural network for reconstruction architecture, which jointly optimizes the sensing and the non-linear reconstruction operators.}
	\centering
	\label{fig:nn_scheme}
\end{figure*}

\begin{table*}[h]
	\centering
	\resizebox{\textwidth}{!}{  
		\begin{tabular}{|c|c|c|c|c|c|c|c|}
			\hline 
			Sensing & No. of & \multicolumn{1}{ |c|  }{\multirow{2}{*}{B} } & \multicolumn{1}{ |c|  }{\multirow{2}{*}{K} } & & Training with SSIM loss & \multicolumn{1}{ |c|  }{\multirow{2}{*}{Training with MSE loss} }  & Training with SSIM loss \\ 
			Rate & Measurements & & & & $W(\signal,\measurement)$ is Equation (\ref{nonUniformWeights}) & & $W(\signal,\measurement) \equiv 1$ \\
			\hline
			\multicolumn{1}{ |c|  }{\multirow{2}{*}{0.125} } &
			\multicolumn{1}{ |c| }{\multirow{2}{*}{128} } &
			\multicolumn{1}{ |c| }{\multirow{2}{*}{2} } &
			\multicolumn{1}{ |c| }{\multirow{2}{*}{1} } &
			SSIM score & 0.885392 & 0.892905 & 0.868903 \\ \cline{5-5} & & & & 
			MSE score & 0.005893 & 0.003517 & 0.005757 \\
			\hline 
			\multicolumn{1}{ |c|  }{\multirow{2}{*}{0.0625} } &
			\multicolumn{1}{ |c| }{\multirow{2}{*}{64} } &
			\multicolumn{1}{ |c| }{\multirow{2}{*}{1} } &
			\multicolumn{1}{ |c| }{\multirow{2}{*}{2} } &
			SSIM score & 0.760333 & 0.793642 & 0.719304 \\ \cline{5-5} & & & & 
			MSE score & 0.010316 & 0.006469 & 0.011553 \\ 
			\hline
		\end{tabular} 
	}
	\vskip 0.3cm
	\caption{SSIM vs. MSE loss score of NN training with SSIM loss function vs. MSE loss function for different Sensing rates}
	\label{Test_results}
\end{table*}

\section{Compressed Sensing Overview}
\label{Compressed Sensing}
Given a signal $\signal \in \mathbf{R}^N$, an $M \times N$ sensing matrix $\Phi$ (such that $M \ll N$) and a measurements vector $\measurement = \Phi \signal$, the goal of CS is to recover the signal from its measurements. The sensing rate is defined by $R=M/N$, and since $R \ll 1$ the recovery of $\signal$ is not possible in the general case. According to CS theory \cite{donoho2006compressed,candes2008introduction}, signals that have a sparse representation in the domain of some linear transform can be exactly recovered with high probability from their measurements: let $\signal = \Psi \reps $, where $\Psi$ is the inverse transform, and $\reps$ is a sparse coefficients vector with only $S \ll N$ non-zeros entries, then the recovered signal is synthesized by $ \hat{\signal} = \Psi \repsHat$, and $\repsHat$ is obtained by solving the following convex optimization problem:
\begin{equation}
\repsHat  = \argmin_{\reps{'}} \left\|\reps{'}\right\|_1 \text{ subject to } \measurement = \Phi \Psi \reps{'},
\end{equation}
where $\left\|\alpha\right\|_1$ is the $l_1$-norm, which is a convex relaxation of the $l_0$ pseudo-norm that counts the number of non-zero entries of $\alpha$. The exact recovery of $\signal$ is guaranteed with high probability if $\reps$ is sufficiently sparse and if certain conditions are met by the sensing matrix and the transform \cite{donoho2006compressed}.

\section{Structural similarity index (SSIM)}
\label{SSIM}
The Structural SIMilarity (SSIM) index \cite{wang2004image} is a method for measuring the similarity between two images. The SSIM index can be viewed as a quality measure of one of the images being compared, provided the other image is regarded as of perfect quality. The difference with respect to other techniques such as Mean Squared Error (MSE) or Peak Signal-to-Noise Ratio (PSNR) is that these approaches estimate absolute errors; on the other hand, SSIM is a perception-based model that considers image degradation as perceived change in structural information, while also incorporating important perceptual phenomena, including both luminance masking and contrast masking terms. Structural information is the idea that the pixels have strong inter-dependencies especially when they are spatially close. These dependencies carry important information about the structure of the objects in the visual scene. Luminance masking is a phenomenon whereby image distortions (in this context) tend to be less visible in bright regions, while contrast masking is a phenomenon whereby distortions become less visible where there is significant activity or "texture" in the image. Given two images \boldmath$X$ and $Y$, let $\signal$ and $\measurement$ be column vector representations of two image patches (e.g., 8 $\times$ 8 windows) extracted from the same spatial location from $X$ and $Y$, respectively. Let $\mu_\signal$, $\sigma_\signal^2$ and $\sigma_{\signal \measurement}$ represent the sample mean of the components of $\signal$, the sample variance of $\signal$ and the sample covariance of $\signal$ and $\measurement$, respectively:
\begin{equation}
\mu_x = \frac{1}{N_P}(\underline{1}^T \cdot \signal)
\label{mu_x}
\end{equation} 
\begin{equation}
\sigma_\signal^2 = \frac{1}{N_P-1}(\signal - \mu_\signal)^T (\signal - \mu_\signal)
\label{sigma_x}
\end{equation}
\begin{equation}
\sigma_{\signal \measurement} = \frac{1}{N_P-1}(\signal - \mu_\signal)^T (\measurement - \mu_\measurement)
\label{sigma_xy}
\end{equation}
where $N_P$ is the number of pixels in patch $\signal$ and \boldmath$1$ is a vector with all entries equaling 1. \\
For two image patches, $\signal$ and $\measurement$ we define
\begin{equation}
\begin{aligned}
\begin{tabular}{c c}
$A_1 = 2\mu_\signal \mu_\measurement + C_1$ &
$A_2 = 2\sigma_{\signal \measurement} + C_2$ \\ \\
$B_1 = \mu_\signal^2 + \mu_\measurement^2 + C_1$ &
$B_2 = \sigma_\signal^2 + \sigma_\measurement^2 + C_2$
\end{tabular}
\end{aligned}
\label{AB}
\end{equation}
 The SSIM index between $\signal$ and $\measurement$ is defined as
\begin{equation}
S(\signal,\measurement) = \frac{A_1 \cdot A_2}{B_1 \cdot B_2}
\end{equation}
where $C_1$ and $C_2$ are small constants. It can be shown that the SSIM index achieves its maximum value of 1 if and only if the two image patches \boldmath$\signal$ and $\measurement$ being compared are exactly the same. \\
The SSIM index is computed using a sliding window approach. The window moves pixel by pixel across the whole image. At each step, the SSIM index is calculated within the local window. Finally, we compute a weighted average of the SSIM indexes to yield an overall SSIM index of the whole image:
\begin{equation}
S(X,Y) = \frac{\sum_{i=1}^{N_S} W(\signal_i,\measurement_i) \cdot S(\signal_i,\measurement_i)}{W(\signal_i,\measurement_i)}
\label{weightedSSIM}
\end{equation}
where \boldmath$\signal_i$ and $\measurement_i$ are the $i$th sampling sliding window in images \boldmath$X$ and $Y$, respectively. \boldmath$N_S$ is the total number of sampling windows. $W(\cdot,\cdot)$ is a weighting function where $W(\signal_i,\measurement_i)$ is the $i$th sampling sliding window weight. \\
If we use uniform weighting function, i.e. $W(\signal_i,\measurement_i) \equiv 1$, equation (\ref{weightedSSIM}) becomes:
\begin{equation}
S(X,Y) = \frac{1}{N_S} \sum_{i=1}^{N_S} S(\signal_i,\measurement_i)
\end{equation}
As developed in appendix \ref{SSIMgradient}, the gradient of SSIM index between \boldmath$X$ and $Y$ with respect to $Y$ is:
\begin{equation}
\begin{aligned}
&\nabla_Y S(X,Y) = \nabla_Y \bigg(\frac{\sum_{i=1}^{N_S} W(\signal_i,\measurement_i) \cdot S(\signal_i,\measurement_i)}{\sum_{i=1}^{N_S} W(\signal_i,\measurement_i)} \bigg) \\
&= \frac{\sum_{i=1}^{N_S} \nabla_Y \bigg(W(\signal_i,\measurement_i) \cdot S(\signal_i,\measurement_i) \bigg)}{\sum_{i=1}^{N_S} W(\signal_i,\measurement_i)} \\
&- \frac{\sum_{i=1}^{N_S} W(\signal_i,\measurement_i) \cdot S(\signal_i,\measurement_i)}{\bigg(\sum_{i=1}^{N_S} W(\signal_i,\measurement_i) \bigg) ^ 2} \cdot \sum_{i=1}^{N_S} \bigg(\nabla_Y W(\signal_i,\measurement_i) \bigg)
\end{aligned}
\label{generalGradient2}
\end{equation}
\normalsize
If we use uniform weighting function, i.e. $W(\signal_i,\measurement_i) \equiv 1$, equation (\ref{generalGradient2}) becomes:
\begin{equation}
\nabla_Y S(X,Y) = \frac{1}{N_S} \sum_{i=1}^{N_S} \nabla_Y S(\signal_i,\measurement_i)
\end{equation}
See appendix \ref{SSIMgradient} and Figure (\ref{fig:gradient_scheme}) for further explanations.

\begin{figure*}[h]
	\includegraphics[width=\linewidth]{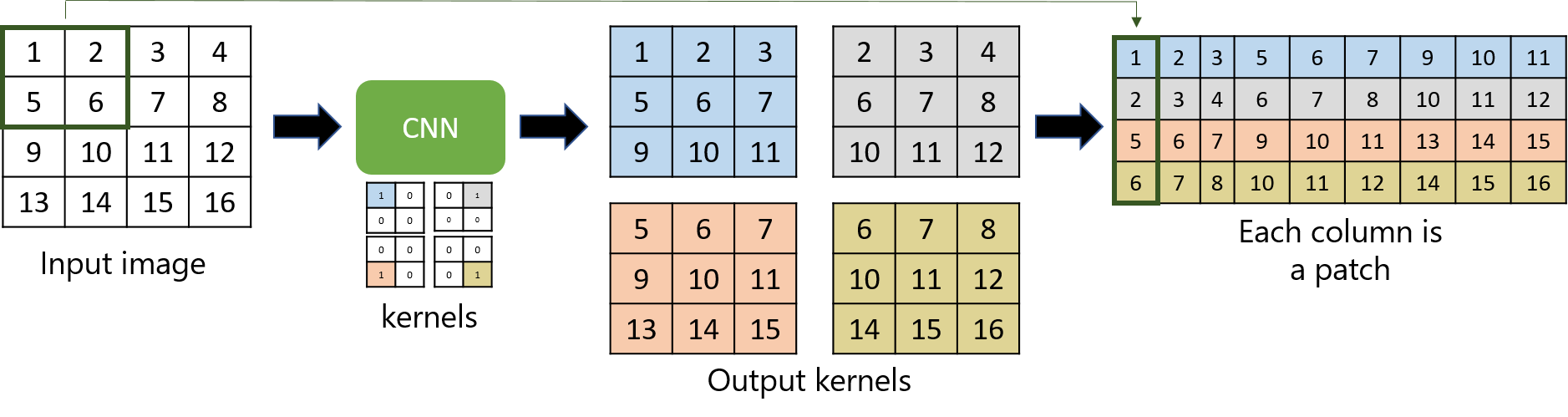}
	\caption{Scheme of the CNN kernels (as described in appendix \ref{im2col}) for extracting 2x2 patches from a 4x4 image. The CNN last layer reshapes the output kernels to 2x2 patches as columns.}
	\centering
	\label{fig:im2col_scheme}
\end{figure*}

\begin{figure*}[h!]
	\includegraphics[width=\linewidth]{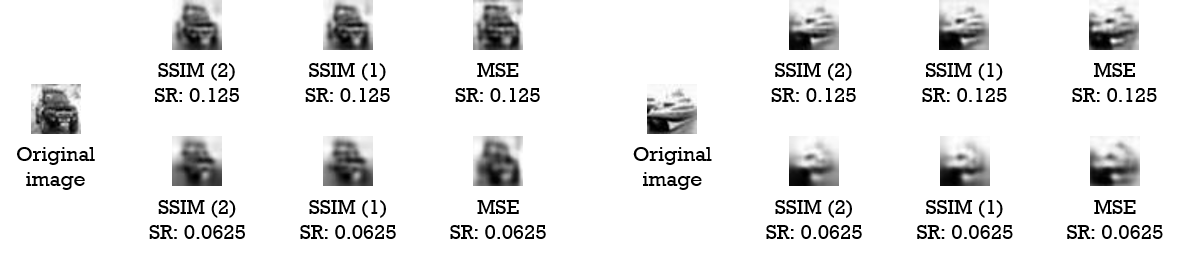}
	\caption{Reconstruction of 2 test images. SSIM(1) is Equation (\ref{nonUniformWeights}) as weight function. SSIM(2) is a uniform weight function $W(\signal,\measurement) \equiv 1$. SR stands for Sensing Rate.}
	\centering
	\label{fig:img_results}
\end{figure*}

\section{The Proposed Approach}
\label{The Proposed Approach}
In this paper we propose an end-to-end deep learning solution for CS, which jointly optimizes the sensing matrix $\Phi$ and the non-linear reconstruction operator, which is parameterized by a coefficients matrix $W$.
The proposed method provides a solution to the following joint optimization problem: 
\begin{equation}
\{\widetilde{\Phi},\widetilde{W}\}= \arg \min_{\Phi,W}{\frac{1}{N}\sum_{i=1}^{N}\mathcal{L}(\mathcal{N}_{W}(\Phi \mathbf{x_{i}}}),\mathbf{x_{i}}),
\end{equation}
where $\{\mathbf{x_{i}}\}_{i=1}^N$ 
is the collection of $N$ signals. The loss function $\mathcal{L}(\bullet,\bullet)  $ measures the distance between the input signal and the reconstructed one, provided by the reconstruction operator $\mathcal{N}_{W}(\bullet)$, whose input is the compressed samples, denoted by  $\Phi \mathbf{x_{i}}$.
In this paper we compare SSIM vs. MSE as loss function, since MSE is commonly used for learning reconstruction networks. Note that during training the sensing layer (matrix) and the subsequent layers, represented  by $\mathcal{N}_{W}(\bullet)$, are treated as a single deep network. However, once training is complete, the sensing matrix is detached from the subsequent reconstruction layers, and used for performing signal sensing. The input of the reconstruction operator, is therefore the second layer of the end-to-end learned network.\\
Our choice is motivated by the success of deep neural networks for the task of full-image reconstruction \cite{hinton2006reducing} in which a neural network based autoencoder achieved state-of-the-art performance. In our approach, the first layer learns the sensing matrix $\widetilde{\Phi}$, and the following fully-connected layers perform the non-linear reconstruction stage. The proposed method was tested on CIFAR10 image recognition database. 

\section{Performance Evaluation}
\label{Results}
This section describes the proposed architectures and provides performance evaluation of the results.
The CIFAR10 \cite{krizhevsky2014cifar} dataset contains 60,000 color images of $32 \times 32$ = 1024 pixels, drawn from 10 different classes. This dataset is divided into training and test sets, containing 50,000 and 10,000 images respectively. Since our proposed network expects grayscale images, for training on CIFAR10 we used only the $1^{st}$ channel for each dataset image. Moreover, we enlarged the training set to 200,000 samples by rotating the original training set images by 90, 180 and 270 degrees. The fully-connected network includes the following layers:
\begin{enumerate}[label=\arabic*.]\setlength\itemsep{0.3em}
	\item An input layer with $N$ nodes.
	\item A compressed sensing fully-connected layer with $NR$ nodes, $R\ll1$ (its weights form the sensing matrix).\vskip -1cm
	\item $K\ge1$ reconstruction layers with $NB$ nodes in each layer, where $B \in \{1,2\}$. Each layer is followed by a sigmoid activation unit.
	\item An output layer with $N$ nodes.
\end{enumerate}
Figure (\ref{fig:nn_scheme}) illustrates the aforementioned flow.
\vskip 0.3cm
The SSIM loss function was used with $8 \times 8$ window size. We tested 2 different weighting functions for the SSIM loss. The first function is the uniform weighting function, i.e. $W(\signal_i,\measurement_i) \equiv 1$. The second function was recommended in \cite{wang2008maximum}:
\begin{equation}
W(\signal,\measurement) = log \bigg[\bigg(1+\frac{\sigma_\signal ^2}{C_2}\bigg) \bigg(1+\frac{\sigma_\measurement ^2}{C_2}\bigg) \bigg]
\label{nonUniformWeights}
\end{equation}
\vskip 0.3cm
The optimization algorithm of choice was Adam \cite{kingma2014adam} with an initial learning rate of $5\cdot10^{-4}$. The training stopping criteria was 50 consecutive epochs without reaching a new minima.
\\Table \ref{Test_results} compares between the neural network trained with SSIM as loss function and the neural network trained with MSE as loss function. The table shows the average SSIM and MSE loss over 10,000 CIFAR10 test images on these networks. The results show that a neural network trained with MSE loss achieves better reconstruction quality even in SSIM index score on CIFAR10 dataset.

\section{Conclusion}
\label{Conclusions}
This paper presents SSIM as loss function for training end-to-end deep neural network for compressed sensing and non-linear reconstruction, in which the sensing matrix and the non-linear reconstruction operator are jointly optimized during the training phase.The proposed approach does not outperform state-of-the-art in terms of reconstruction quality by SSIM loss. Since calculating SSIM score is more complicated than calculating MSE score for a given image, as explained in details on appendix \ref{im2col}, our approach can be further improved by combining the SSIM as loss function with block-based compressed sensing approach \cite{adler2016deep}. Learning a deep neural network for blocks reconstruction under the SSIM loss would be significantly faster than learning a network for full-image reconstruction in cases where there are many patches per image. Another possible future work direction is to expand the SSIM loss function to multi-scale SSIM loss function as described on \cite{zhao2017loss}.


\appendices
\section{Efficient training with SSIM loss}
\label{im2col}
As described on section \ref{SSIM}, using SSIM as a loss function for reconstruction neural network, instead of the commonly used MSE function, sets few computational challenges since the SSIM function is much more complicated computationally compared to the MSE function. \\
The $1^{st}$ challenge is to divide the image to patches. Extracting all image patches might be time consuming if it is programmed straightforward, i.e. using for-loop for example. We experienced significant acceleration once we replaced the for-loop with another method. Our efficient method for extracting the image patches was to build a convolutional neural-network (CNN). The CNN is built from kernels with the same dimensions as the SSIM window dimensions. Each kernel has 0's all over and a value of 1 in a specific position, i.e. each kernel has the value of 1 in a different position. When we forward an image through the CNN it outputs kernels where each kernel holds all the possible values from the original image that should be placed in the kernel position as part of some patch. The CNN final layer reshapes the kernels to the patches shape as required. See Figure (\ref{fig:im2col_scheme}) for further explanations. \\
Another challenge is the trade-off between computation time and storage size. In each training epoch, we calculate SSIM loss for a batch of samples from the training set. It means we have to calculate SSIM loss for the same batch more than once during training, since we iterate over the training set more than once. On one hand, we can calculate once equations (\ref{mu_x}),(\ref{sigma_x}) prior to training and store the results in memory. It saves us computations during training. On the other hand, it requires some storage space. We can extend the dilemma even to storage of training set patches. We can trade-off between extracting the patches in each epoch, which costs us in computation time, to extracting all training set patches pre-training and store them in memory, which costs us in significant storage space.

\begin{figure*}[h]
	\centering
	\includegraphics[width=\columnwidth,height=5cm]{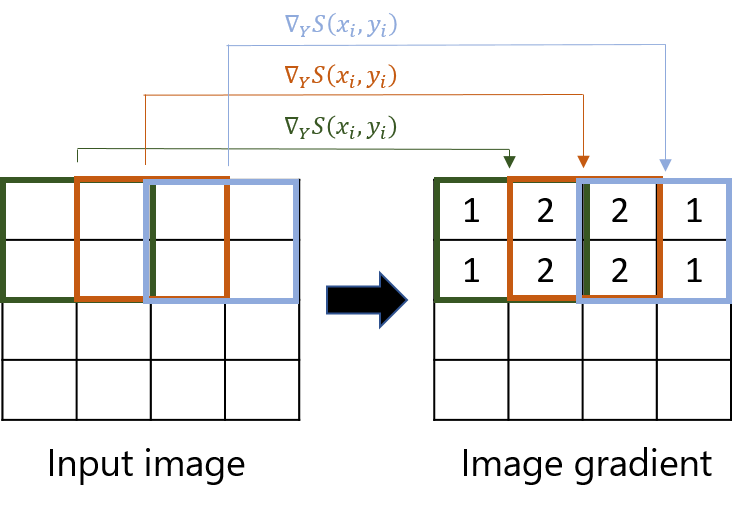}
	\caption{Scheme of calculating overall SSIM gradient for whole image. The overall SSIM gradient is calculated per pixel. For each pixel we sum all the gradients of windows that include it and divide by the total number of gradients. The numbers inside the image gradient pixels represents the number of gradients we sum for each pixel, i.e. the pixels in the borders sum less gradients than the central pixels.}
	\centering
	\label{fig:gradient_scheme}
\end{figure*}

\section{SSIM Index Gradient}
\label{SSIMgradient}
Given an SSIM index:
\begin{equation}
S(X,Y) = \frac{\sum_{i=1}^{N_S} W(\signal_i,\measurement_i) \cdot S(\signal_i,\measurement_i)}{\sum_{i=1}^{N_S} W(\signal_i,\measurement_i)}
\end{equation}
We are interested in its gradient with respect to $Y$, i.e. $\nabla_Y S(X,Y)$.
\\ Let us start with a simple case, where there is single window and a uniform weighting function $W(\signal_i,\measurement_i) \equiv 1$, therefore:
\begin{equation}
S(X,Y) = S(\signal_1,\measurement_1)
\end{equation}
Equations (\ref{mu_x}),(\ref{sigma_x}),(\ref{sigma_xy}) and (\ref{AB}) induce the following equivalence:
\begin{equation}
S(X,Y) = S(\signal_1,\measurement_1) = \frac{A_1 \cdot A_2}{B_1 \cdot B_2}
\end{equation}
Let us develop the gradient with respect to $Y$:
\begin{equation}
\nabla_Y A_1 = 2 \mu_\signal (\nabla_Y \mu_\measurement) =  2 \mu_\signal \bigg(\frac{1}{N_P} \cdot \underline{1} \bigg)
\end{equation}
\begin{equation}
\begin{aligned}
\nabla_Y B_1 &= \nabla_Y (\mu_\measurement ^ 2) = 2 \mu_\measurement (\nabla_Y \mu_\measurement) = \\
&= 2 \mu_\measurement \bigg(\frac{1}{N_P} \cdot \underline{1} \bigg)
\end{aligned}
\end{equation}
\begin{equation}
\nabla_Y A_2 = 2 (\nabla_Y \sigma_{\signal \measurement}) =  2 \bigg(\frac{1}{N_P - 1} \cdot (\signal - \mu_\signal) \bigg)
\end{equation}
\begin{equation}
\nabla_Y B_2 = \nabla_Y (\sigma_\measurement ^ 2) =  \frac{2}{N_P - 1} (\measurement - \mu_\measurement)
\end{equation}
\begin{equation}
\nabla_Y (A_1 A_2) = \big(\nabla_Y A_1 \big)A_2 + A_1 \big(\nabla_Y A_2 \big)
\end{equation}
\begin{equation}
\nabla_Y (B_1 B_2) = \big(\nabla_Y B_1 \big)B_2 + B_1 \big(\nabla_Y B_2 \big)
\end{equation}
Therefore, the SSIM index gradient with respect to $Y$:
\begin{equation}
\begin{aligned}
&\nabla_Y S(\signal_1,\measurement_1) = \\
\frac{\nabla_Y (A_1 A_2)}{(B_1 B_2)} &- \frac{(A_1 A_2)}{(B_1 B_2) ^ 2} \cdot \bigg(\nabla_Y (B_1 B_2)\bigg)
\end{aligned}
\label{singleWindowGradient}
\end{equation}
Let us develop $\nabla_Y S(X,Y)$ for the general case:
\small
\begin{equation}
\begin{aligned}
\nabla_Y  \bigg(W(\signal_i,\measurement_i) &\cdot S(\signal_i,\measurement_i) \bigg) = \\
\bigg(\nabla_Y W(\signal_i,\measurement_i) \bigg) \cdot S(\signal_i,\measurement_i)
&+ W(\signal_i,\measurement_i) \cdot \bigg(\nabla_Y S(\signal_i,\measurement_i) \bigg)
\end{aligned}
\end{equation}
\begin{equation}
\begin{aligned}
&\nabla_Y S(X,Y) = \nabla_Y \bigg(\frac{\sum_{i=1}^{N_S} W(\signal_i,\measurement_i) \cdot S(\signal_i,\measurement_i)}{\sum_{i=1}^{N_S} W(\signal_i,\measurement_i)} \bigg) \\
&= \frac{\sum_{i=1}^{N_S} \nabla_Y \bigg(W(\signal_i,\measurement_i) \cdot S(\signal_i,\measurement_i) \bigg)}{\sum_{i=1}^{N_S} W(\signal_i,\measurement_i)} \\
&- \frac{\sum_{i=1}^{N_S} W(\signal_i,\measurement_i) \cdot S(\signal_i,\measurement_i)}{\bigg(\sum_{i=1}^{N_S} W(\signal_i,\measurement_i) \bigg) ^ 2} \cdot \sum_{i=1}^{N_S} \bigg(\nabla_Y W(\signal_i,\measurement_i) \bigg)
\end{aligned}
\label{generalGradient}
\end{equation}
\normalsize
If we use uniform weighting function, i.e. $W(\signal_i,\measurement_i) \equiv 1$, equation (\ref{generalGradient}) becomes:
\begin{equation}
\nabla_Y S(X,Y) = \frac{1}{N_S} \sum_{i=1}^{N_S} \nabla_Y S(\signal_i,\measurement_i)
\label{uniformWeightsGradient}
\end{equation}
Let us recall that $X$ and $Y$ are images, \boldmath$\signal$ and $\measurement$ are windows extracted from $X$ and $Y$ respectively, therefore the gradient $\nabla_Y S(\signal,\measurement)$ has same dimensions as $\signal$ and $\measurement$. Notice the sum in (\ref{uniformWeightsGradient}) \underline{is not} a standard sum. We can think about it as we have an image with the same dimensions as $X$ and $Y$ which all its values are 0, then we sum each gradient $\nabla_Y S(\signal_i,\measurement_i)$ to its window location and finally divide the whole image by $N_S$. The same follows for sums in the general gradient equation (\ref{generalGradient}), each sum is not a standard sum, but means we should sum each gradient to its window location. See Figure (\ref{fig:gradient_scheme}) for further explanations.


\ifCLASSOPTIONcaptionsoff
  \newpage
\fi



%

\bibliography{refs}{}
\bibliographystyle{IEEEtran}

%




\end{document}